\documentclass[12pt]{article}

\usepackage{newtxtext,newtxmath}

\usepackage{graphicx}

\usepackage[letterpaper,margin=1in]{geometry}

\renewenvironment{abstract}
	{\quotation}
	{\endquotation}

\date{}


\usepackage{scicite}

\usepackage{url}

\usepackage{times}
\usepackage{graphicx}
\usepackage{caption}
\usepackage{tabularx}
\usepackage{booktabs}
\usepackage{makecell}
\usepackage{multirow}
\usepackage{xcolor}
\usepackage{bm}
\usepackage{lineno}

\newcommand{\model}{\mbox{ORCA-DL}}
\newcommand{\nino}{Niño}
\newcommand{\nina}{Niña}
\newcommand{\du}{\textdegree}

\def\scititle{
	Data-driven Global Ocean Modeling for Seasonal to Decadal Prediction
}

\title{\bfseries \boldmath \scititle}

\author{
	Zijie~Guo$^{1,2}$,
	Pumeng~Lyu$^{2}$,
	Fenghua~Ling$^{2}$,\and
        Lei~Bai$^{2\ast}$,
        Jing-Jia~Luo$^{3\ast}$,
        Niklas~Boers$^{4,5}$,
        Toshio~Yamagata$^{6}$, \and
        Takeshi~Izumo$^{7}$,
        Sophie~Cravatte$^{8,9}$,
        Antonietta~Capotondi$^{10,11}$,
        Wanli~Ouyang$^{2}$\and
	\small$^{1}$School of Computer Science, Fudan University, Shanghai, China.\and
	\small$^{2}$Shanghai Artificial Intelligence Laboratory, Shanghai, China.\and
 	\small$^{3}$Institute for Climate and Application Research (ICAR)/School of Future Technology,\and \small Nanjing University of Information Science and Technology, Nanjing, China.\and
  	\small$^{4}$Earth System Modelling, School of Engineering and Design, \and \small Technical University of Munich, Munich, Germany.  \and 
   \small$^{5}$ Potsdam Institute for Climate Impact Research, Potsdam, Germany. \and
   \small$^{6}$ Application Laboratory, Japan Agency for Marine-Earth Science and Technology, Yokohama, Japan. \and
   \small$^{7}$ Institut de Recherche pour le Développement (IRD),  UMR241 SECOPOL (ex-EIO), Tahiti, French Polynesia \and
   \small$^{8}$ LEGOS (IRD/UT3/CNES/CNRS), Université de Toulous, Toulouse, France. \and
   \small$^{9}$ French National Research Institute for Sustained Development (IRD), Nouméa, New Caledonia \and
   \small$^{10}$ Cooperative Institute for Research in Environmental Sciences,\and \small University of Colorado Boulder, Boulder, Colorado, America. \and
   \small$^{11}$ NOAA Physical Sciences Laboratory, Boulder, Colorado, America. \\
	\small$^\ast$Corresponding authors. Email: bailei@pjlab.org.cn; luojj@nuist.edu.cn\and
}

\begin{document} 

\renewcommand\linenumberfont{\normalfont\small}
\setlength\linenumbersep{25pt}
\maketitle

\begin{abstract} \bfseries \boldmath

Accurate ocean dynamics modeling is crucial for enhancing understanding of ocean circulation, predicting climate variability, and tackling challenges posed by climate change. Despite improvements in traditional numerical models, predicting global ocean variability over multi-year scales remains challenging. Here, we propose ORCA-DL (Oceanic Reliable foreCAst via Deep Learning), the first data-driven 3D ocean model for seasonal to decadal prediction of global ocean circulation. ORCA-DL accurately simulates three-dimensional ocean dynamics and outperforms state-of-the-art dynamical models in capturing extreme events, including El \nino-Southern Oscillation and upper ocean heatwaves. This demonstrates the high potential of data-driven models for efficient and accurate global ocean forecasting. Moreover, ORCA-DL stably emulates ocean dynamics at decadal timescales, demonstrating its potential even for skillful decadal predictions and climate projections.

\end{abstract}

\noindent
Climate predictions from seasonal to decadal timescales (S2D) are essential for society, as they support resource management, planning, and long-term infrastructure investments in response to climate variability and change~\cite{troccoli2010seasonal,meehl2021initialized}. Despite significant international efforts over the past decade to enhance prediction skills~\cite{randall2000general}, the accuracy of climate predictions is still lacking. In addition, the reliability of climate model projections is also often hampered by model deficiencies
~\cite{boberg2012overestimation}. 
Encouragingly, a body of research indicates that this challenge can be addressed by improving the modeling of more predictable subsystems within the climate system, particularly ocean modeling
~\cite{shukla2000dynamical,kanamitsu2002ncep}, since ocean dynamics is crucial in controlling weather patterns and driving climate variability~\cite{wang2019three,hasselmann1982ocean,kirtman2013ocean}.

Traditionally, Ocean General Circulation Models (OGCMs) have been employed to simulate the time-evolving, three-dimensional dynamics of ocean currents, temperature, and salinity~\cite{griffies2000developments}. And they have then been coupled to an atmospheric GCM (AGCM) to obtain a Coupled General Circulation Model (CGCM) that is able to simulate and forecast climate and its modes of variability such as El \nino-Southern Oscillation (ENSO). While these models possess a robust physical framework, they often depend on physical parameterization due to computational limitations~\cite{palmer2005representing}. This reliance can restrict the accuracy of model simulations and climate predictions, as certain physical processes may not be fully captured. Additionally, the slow development of dynamical models does not fully exploit the rapidly increasing amounts of available observational data~\cite{edwards2011history,schneider2023harnessing}. Consequently, statistical and artificial intelligence (AI) methods are becoming increasingly popular to achieve more accurate ocean predictions~\cite{ham2019deep,ling2022multi,qin2022deep,subel2024building}. Existing AI studies primarily focused on forecasting specific climate phenomena such as the ENSO~\cite{ham2019deep}, the Indian Ocean Dipole (IOD)~\cite{ling2022multi}, the Pacific Decadal Oscillation (PDO)~\cite{qin2022deep}, or modeling the three-dimensional ocean structure in specific regions~\cite{subel2024building}. While these innovative machine-learning approaches hold transformative potential, they are still far from achieving reliable simulations and predictions of the global three-dimensional ocean.

Here, to tackle this challenge, we present \model~(\textbf{O}ceanic \textbf{R}eliable fore\textbf{CA}st via \textbf{D}eep \textbf{L}earning), the first data-driven model for seasonal to decadal global ocean circulation predictions. 
Using reanalysis data over the past four decades as a predictand, \model~outperforms state-of-the-art dynamical models and demonstrates skillful 
predictions across three dimensions and multiple key variables, including salinity, potential temperature, and currents. \model~produces 
physically consistent forecasts for various climate phenomena, including ENSO and marine heatwaves, 
and can be stably run for decadally-long simulations. These results highlight the ability of \model's to propel global ocean modeling, as well as to enhance the computational efficiency of climate predictions.

\section*{Results}

\model~utilizes the initial conditions of the present global ocean states, including variables like temperature, salinity, and currents, initially forced by the atmosphere through solely two variables, zonal and meridional wind stress (see table~\ref{tab:vars} for a complete list of modeling variables), to generate a month-by-month forecast of the future ocean state. Similar to the dynamical models, \model~is capable of autoregressively generating predictions of arbitrary length. 
Since the monthly reanalysis data are scarce, to increase the number of training data and reduce the systematic biases among dynamical models, historical simulations from 20 models (table~\ref{tab:cmip6}) that participated in the Coupled Model Intercomparison
Project phase 6 (CMIP6) \cite{eyring2016overview} are used for training \model. And the data before 1980 from reanalysis datasets Simple Ocean Data Assimilation (SODA) \cite{carton2000simple} and Ocean Reanalysis System 5 (ORAS5) \cite{copernicus2021oras5} are used for parameters tuning during the model development. After training, data from the NCEP Global Ocean Data Assimilation System (GODAS) \cite{behringer1998improved} is used as observational ground truth to evaluate the skill independently.

\model~mainly consists of several encoders, decoders, and a fusion module. Inspired by traditional ocean modeling frameworks, \model~separately encodes each physical ocean variable while incorporating coastlines as boundary conditions. High-dimensional encoded signals of each ocean variable are integrated in the fusion module with the encoded surface forcing condition (wind stresses) to simulate the complex operations of the dynamical equations. Finally, the decoders restore the calculated results to the spatial domain of each ocean variable, yielding next-step predictions.
The overall architecture and process of making a forecast are illustrated in fig.~\ref{fig:model} and fig.~\ref{fig:rollout_example} respectively.

Since a comprehensive long-term ocean forecast requires coupling ocean models with atmospheric signals, we embed atmospheric forcing with varying lead times, allowing the model to account for the decaying influence of initial atmospheric conditions on ocean dynamics over time.
To further improve the accuracy and efficiency of long-term predictions, \model~assigns intermediate hidden features across different lead times, adjusting its forecasts accordingly. Additionally, a 10-member ensemble modeling is used to make more reliable predictions that account for inherent uncertainties. The results shown in the next sections, unless otherwise specified, are based on ensemble mean predictions (See section ``\model~Model" in Supplementary Materials and Methods).



\subsection*{Physical consistency of the mean state from \model's simulation}
Dynamical models often exhibit systematic biases, and their simulated climate states are still significantly
different from the real world~\cite{doblas2013seasonal,magnusson2013evaluation}. Therefore, evaluating the model's climate state characteristics is crucial before assessing its skill in predicting climate variations (Fig.~\ref{fig:tem_sal_vel_clim} and figs.~\ref{fig:supp_temp_current_clim} to~\ref{fig:supp_temp_salt_clim}). 

As illustrated in Fig.~\ref{fig:tem_sal_vel_clim} A-B, even though \model~is trained on model-simulated data, it still demonstrates strong ability to simulate global sea surface temperature (SST) and sea surface current (SSC) patterns up to 12 months ahead, allowing accurate climate features to be obtained. In contrast to the pronounced warm bias exhibited by the operational dynamical model (NUIST-CFS1.2~\cite{huihang41impact}, see section ``Data" in Supplementary Materials and Methods), \model~presents a significantly reduced overall bias, although it tends to underestimate equatorial temperatures (Fig.~\ref{fig:tem_sal_vel_clim} C-D and fig.~\ref{fig:supp_temp_clim_bias}). Additionally, \model~successfully captures several major ocean currents, including the Antarctic Circumpolar Current, the Indian Ocean Gyre, and the North and South Equatorial Currents. The skillful simulation of these currents is promising for accurately predicting climate phenomena that depend on ocean advection.

The scatter plots in Fig.~\ref{fig:tem_sal_vel_clim} E-H further demonstrate the consistency of the \model~predictions of surface temperature and salinity. Although there is a small divergence in the potential density ($\sigma_0$) range of  24 to 26 $\rm kg/m^3$ (dashed contour lines in Fig.~\ref{fig:tem_sal_vel_clim} E-H), \model~maintains a higher correlation coefficient compared to NUIST-CFS1.2 (fig.~\ref{fig:supp_temp_salt_clim} e-h), which is calculated based on the density distribution of points as shown in fig.~\ref{fig:supp_tem_sal_occurrence}. 
Since the temperature and salinity values of water masses in different regions are different, we further visualized these values for different water masses in fig.~\ref{fig:supp_water_mass}, which demonstrates that \model~can skillfully models the thermohaline properties at different depths and in different regions.

These results demonstrate that \model~can accurately capture the 3D oceanic state, in terms of density and currents, and the physical consistency among different variables.
While \model~effectively simulates the climate state, it is not exempt from systematic deviations, such as climate drift~\cite{doblas2013seasonal,magnusson2013evaluation}.
However, it is noteworthy that \model's climate drift is substantially smaller compared to other dynamical models, regardless of any variable (fig.~\ref{fig:supp_clim_drift}).

\subsection*{Skillful Seasonal to Interannual Forecasting Skills of \model}


\subsubsection*{Overall performance of \model}
To further assess the prediction skill of \model, we show the global geographical distribution of the temporal anomaly correlation coefficient (ACC) for temperature and salinity at the surface and at a depth of 300m for different lead times in Fig.~\ref{fig:temp_salt_acc_pattern} (see fig.~\ref{fig:supp_current_acc_pattern} for current prediction). For SST, \model's highest skill appears in the tropics and especially in the tropical Pacific, where SST anomalies are the strongest and the most persistent and where ENSO brings a relatively high predictability even at lead times up to 18 months (Fig.~\ref{fig:temp_salt_acc_pattern} A-C). 
For other variables, whether considering temperature and salinity changes in the surface layer or in the deep layer, \model~demonstrates prediction skills superior or comparable to those of the traditional numerical models in the equatorial zone and Northern Hemisphere (cf. Fig.~\ref{fig:temp_salt_acc_pattern} and figs.~\ref{fig:supp_current_acc_pattern} to~\ref{fig:supp_nuist_acc_pattern}). 
Notably, \model~exhibits exceptionally higher performance in the mid and high latitudes of the Southern Hemisphere. This may be attributed to the greater variance of ocean features in the extratropics at deeper layers (fig.~\ref{fig:supp_godas_std}), which causes larger training loss for the AI model. Such regions may be optimized first~\cite{chen2018gradnorm}, similar to more accurate predictions of SST in the tropical Pacific region (Fig.~\ref{fig:temp_salt_acc_pattern} A-C and fig.~\ref{fig:supp_godas_std} a). 

To further highlight the performance of \model, we compare the forecast skills with traditional numerical circulation models, including NUIST-CFS1.2, SINTEX-F~\cite{luo2005seasonal}, and seven models of the North American MultiModel Ensemble (NMME) \cite{kirtman2014north} (figs.~\ref{fig:supp_rmse_acc} to~\ref{fig:supp_vertical_acc}). Both the ACC skills and root mean square error (RMSE) in predicting the different oceanic variables by \model~are superior to most of the traditional dynamical counterparts. Notably, while persistence forecasts demonstrate high skill for temperature prediction in the upper ocean, \model~exhibits a significant improvement over these persistence forecasts (fig.~\ref{fig:supp_vertical_rmse} a-d and fig.~\ref{fig:supp_vertical_acc} a-d).

\subsubsection*{Performance in predicting extreme climate phenomena}

In addition to the overall prediction skills as reported above, we assess \model's capability to predict several key large-scale extreme climatic phenomena.

\noindent \textbf{El \nino-Southern Oscillation (ENSO)} is one of the most prominent climate modes, centered in the central and eastern tropical Pacific Ocean. With global teleconnections, ENSO causes extreme atmospheric and oceanic anomalies worldwide~\cite{bjerknes1969atmospheric,timmermann2018nino}. Predictive skill for this interannual mode has been pushed from seasonal to interannual times scales by deep learning approaches~\cite{ham2019deep,lyu2024resonet}. As a crucial area for monitoring canonical ENSO behavior, the \nino~3.4 region (170\du-120\du W, 5\du S-5\du N) has been extensively used for ENSO studies~\cite{trenberth1997definition,cane2005evolution}. 
The \nino~3.4 index during the December-January-February (DJF) season reflects ENSO amplitude during its peak period. Hindcasts of the \nino~3.4 index during the past four decades at different lead times indicate that \model~can correctly predict the ENSO  \nino~3.4 index even at a long lead time (Fig.~\ref{fig:nino_pattern_index} A). Although predicting ENSO becomes increasingly challenging with longer lead times, \model~still has a highly significant correlation skill of 0.75 at 12-month lead through the spring predictability barrier, and can still well predict the peaks of some extreme ENSO events, such as in the La \nina~events of 1988/89 and 1999/2000, partly predicted even at 24-month lead, and the 2015/16 El \nino, predicted at a 12-month lead.

To put \model's prediction skill for ENSO into context, we compare it to the prediction skill of several state-of-the-art process-based numerical models (Fig.~\ref{fig:nino_pattern_index} B) as well as other two well-known data-driven models (fig.~\ref{fig:supp_ai_enso}). \model's prediction skill for the \nino~3.4 index surpasses that of not only the dynamical models but also data-driven models specifically designed for ENSO prediction, especially for leads longer than 10 months, with a skillful prediction of the \nino~3.4 index extending up to 20 months lead. Additionally, \model, similarly to ENSO-specific AI models~\cite{ham2019deep,zhou2023self}, effectively captures the signal during ENSO phase transitions and is less affected by the spring predictability barrier (fig.~\ref{fig:supp_ai_enso} b-c). 

Further analysis of the temporal evolution of the \nino~3.4 area-averaged potential temperature anomalies, from the surface to a depth of 500 meters (fig.~\ref{fig:supp_nino_ano_pattern}), 
clarifies why \model~achieves such accurate ENSO predictions. The subsurface anomalies exhibit an alternating pattern of warm and cold phases along the thermocline (i.e. the layer with the strongest vertical stratification), in association with the deepening and shallowing of the thermocline. They are related to equatorial Kelvin and Rossby waves propagating eastward and westward respectively, which are crucial for the development and termination of El \nino~and La \nina~events~\cite{izumo2024hybrid}, and seem to be partly captured by \model, at least at 6 and 12-month leads (Fig.~\ref{fig:nino_pattern_index} C-D and fig.~\ref{fig:supp_d20}). 
These subsurface/thermocline depth signals often provide earlier indications of ENSO events and serve as a reliable source for prediction~\cite{guilderson1998abrupt}. 


Additionally, Fig.~\ref{fig:nino_pattern_index} E-G illustrate the depth-longitude distribution of ACC skills in predicting upper ocean temperature in the equatorial region, reinforcing \model's high performance. \model~demonstrates relatively high ACC skills at layers both above 200 meters and below 500 meters, particularly in the surface layer of the central Pacific, consistent with Fig.~\ref{fig:temp_salt_acc_pattern} A-F. Conversely, traditional dynamical models encounter significant forecasting difficulties at depths around 300 meters, nearly losing predictive skill in deeper layers (fig. \ref{fig:supp_nino_ano_pattern} f-i). 

Overall, \model~exhibits excellent three-dimensional modeling capabilities.
Accordingly, \model~has also better skill than the other models in predicting the Central Pacific (CP) \nino~4, and for Eastern Pacific (EP) \nino~3 indices, suggesting that \model~also better forecasts ENSO spatial diversity/flavors and related ENSO continuum between CP El \nino/Modoki events and extreme EP El \nino~events~\cite{ashok2007nino,takahashi2011enso,capotondi2015understanding}. \model's strong predictive performance extends beyond ENSO. It also shows acceptable skills in forecasting other significant climate phenomena, such as the Indian Ocean Dipole (IOD) and the Atlantic \nino~(fig.~\ref{fig:supp_indices}).

\noindent \textbf{Marine heatwaves (MHWs)}, are other oceanic extreme events, characterized by high ocean temperatures which can last from weeks to even years, with huge ecological and socio-economic impacts~\cite{smith2021socioeconomic,santora2020habitat}.
While there is growing recognition of the need to improve upper ocean MHWs forecasts~\cite{smith2021socioeconomic,marin2022local}, most previous studies still focus on surface MHWs due to limitations in ocean data and numerical models' performance~\cite{oliver2021marine,jacox2022global}. 
Based on the aforementioned assessments, we find that \model~exhibits superior multi-layer forecasting capabilities. Consequently, we define upper ocean MHWs based on monthly ocean heat content above 300m depth~\cite{jacox2022global,mcadam2023seasonal} and validate \model's capabilities for forecasting these MHWs. In simple terms, if the anomaly at a location exceeds the 90th percentile of anomalies for that location, it is considered that a heat wave event has occurred there. See section ``Definition of Upper Ocean MHWs" in Supplementary Materials for details.

Fig.~\ref{fig:upper_ocean_mhw} A shows the frequency distribution of upper ocean MHWs calculated using GODAS data during 2010-2018 (the 90th percentile is calculated based on the period of 1985-2018). We find a clear difference in the distribution of frequency of upper ocean MHWs and surface MHWs over the past decade as shown in fig.~\ref{fig:supp_surface_mhw} a-b. 
The frequency of surface MHWs in some areas, such as near French Polynesia—which encompasses about half of all the world's atolls—is significantly underestimated, making it particularly vulnerable to MHWs that can lead to coral bleaching.
Thus, the forecast of upper ocean MHWs cannot be ignored.


In terms of predictive ability, \model~performs exceptionally well in areas with a high frequency of occurrence of upper ocean MHWs (see table~\ref{tab:mhw_region} for specific locations), surpassing SINTEX-F, NUIST-CFS1.2, and persistence forecasts. It is noteworthy that traditional dynamical models struggle to exceed the accuracy of persistence forecasts. Compared with SINTEX-F, NUIST-CFS1.2 achieves comparable performance to \model~in some regions, possibly thanks to subsurface ocean data assimilation. 
Accordingly, we show the comparison of the forecast performance of surface MHWs in fig.~\ref{fig:supp_surface_mhw}. Nevertheless, \model~either demonstrates a clear advantage or performs comparably well relative to all physical models.
This further highlights the \model's capability to accurately predict extreme events in both the surface and upper ocean.


\subsection*{Potential of Decadal Prediction}

Having explored the performance of \model~in seasonal and interannual forecasts, we extend our evaluation to decadal predictions, which provide a crucial foundation for informing adaptation strategies as the climate evolves~\cite{meehl2009decadal}.
While decadal predictions typically account for external influences, such as carbon dioxide emissions, our current focus is on assessing the model's ability to produce stable long-term simulations and accurately reproduce climatology, seasonal cycles, and other variability modes internal to the climate system, including those occurring at decadal timescales. This evaluation does not involve comparisons with traditional methods and future work will aim to develop more advanced deep learning-based models for decadal forecasts that incorporate changes in external forcing.

Given that rolling forecast over such long periods poses challenges for an AI model due to error accumulation~\cite{bi2022pangu}, we first verify that the model can run stably on the decadal time scale. Fig.~\ref{fig:decadal_prediction} A illustrates a 10-year variation of the predicted global mean SST initiated from December 1989. Given the initial condition in December 1989, \model~autoregressively predicts the ocean states from January 1990 to December 1999 (see fig.~\ref{fig:supp_decadal_diff_init} for prediction initialized from different years).
As the lead months increase, \model~gradually exhibits a slight cold bias. Nevertheless, there is remarkable consistency between the predictions of \model~and GODAS observations evidenced by a correlation skill of 0.77 and this stability is largely independent of the initial conditions (fig.~\ref{fig:supp_decadal_diff_init}).
Typically, AI models based on autoregressive forecasting tend to accumulate errors quickly, leading to significant bias in forecast results. In contrast, our model maintains only a small cold bias. 
Additionally, we note that \model~can still capture the seasonal cycles in SST without solar radiation as an input. One possible reason is that \model~uses lead time as an input, and since SST exhibits significant seasonal variations, the model can implicitly learn the annual cycles during training.

Additionally, based on the decadal prediction of \model, we observed significant cold biases in the decadal mean-states of SST in some regions (fig.~\ref{fig:supp_decadal_bias_depth} a), such as the Kuroshio Extension region, the Atlantic Ocean, etc.,
which results in the slightly colder global mean SST (Fig.~\ref{fig:decadal_prediction} A). Furthermore, the surface current velocity along the equator is notably underestimated (fig.~\ref{fig:supp_decadal_bias_depth} c). Interestingly, these biases in \model~without $\rm CO_2$ input are roughly consistent with past climate changes in response to increased $\rm CO_2$ forcing in many regions as shown in fig.~\ref{fig:supp_godas_trend}. This consistency is also preserved at deeper layers (fig.~\ref{fig:supp_decadal_bias_depth} d-o). This indicates the potential of \model~to enhance the reliability of decadal forecasts by incorporating realistic $\rm CO_2$ and other radiative forcing in the future.

To further explore the model's ability to capture decadal variations, the prediction of global mean SST anomalies under different lead years is shown in Fig.~\ref{fig:decadal_prediction} B. As we can observe, although the correlation skills decrease at longer lead times, such as five years, it still effectively captures the overall trend. The skills for forecasting the indices of the Pacific Decadal Oscillation (PDO), Interdecadal Pacific Oscillation (IPO), and Atlantic Multidecadal Oscillation (AMO) from seasonal to decadal time-scale, as shown in Fig.~\ref{fig:decadal_prediction} C-D, further demonstrate that \model~exhibits excellent performance in capturing these important climate variations at a relatively short time scale of one to two years and possess potential for longer-term forecasts, such as decadal.

\section*{Discussion}

In this paper, we present \model, the first data-driven global ocean forecasting model for seasonal to decadal predictions. By training with historical simulations from 20 models in CMIP6, \model~exhibits an extraordinary forecast skill almost for all targeted variables, except for slightly inferior ACC skills for the salinity when the lead time exceeds eleven months (one possible reason is that salinity observations are scarce and exhibits relatively small variations compared to other variables, making the persistence forecast a strong baseline). 

As compared to conventional CGCMs, some really interesting improvements are seen in the tropical Pacific, for ENSO, and in the Southern Ocean, with forecasting skills being significantly improved at 12 to 24 months lead. The skillful predictions of ENSO and upper ocean MHWs demonstrate the ability of \model~to detect and predict key climate oscillations and extreme events. The performance of \model~in capturing subsurface dynamics underscores its potential for accurately predicting and capturing subsurface variations. Furthermore, a decadal forecast of global SST with high spatial-temporal consistency and a relatively small bias demonstrates \model's long-term modeling capacities of the ocean. In particular, this result shows that \model~can run stably over a decade, making it a promising candidate also for future climate projections.

Due to insufficient observational data at monthly resolution, we had to train \model~using the simulated data from a large ensemble of dynamical climate models that exhibit biases. Therefore, there is still room for further development toward more effective transfer learning approaches, tailored to the specific requirements of such a large model. Furthermore, adding surface turbulent heat fluxes, radiative fluxes and even other atmospheric variables such as precipitation, surface air temperature, and specific humidity would be the next step to further improve global ocean prediction, as well as improving climate and atmosphere forecasts. Another promising perspective of this \model~is that one could use it to perform sensitivity experiments with oceanic states only in limited regions and see which one matters the most for prediction (as done in~\cite{zhao2024explainable}, in a much simpler model).
This would be possible computationally, thanks to the cost-efficiency of running the model after it has been trained. In addition, we are particularly interested in incorporating external constraints such as carbon dioxide concentrations and physical laws into deep learning models in our future studies.



\begin{figure}[p]
    \centering
    \includegraphics[width=0.99\textwidth]{pdf/tem_sal_vel_clim.pdf}
    \caption{\textbf{Climatological mean states of 1985-2018.} (\textbf{A} to \textbf{B}) The mean states of sea surface temperature (SST; shading) and sea surface currents (SSC; vector) in the June-July-August (JJA) and December-January-February (DJF) seasons based on the predictions from \model~at a lead time of 12 months. (\textbf{C} to \textbf{D}) The bias of SST mean-states in the \model's predictions at 12 months lead relative to GODAS in the JJA and DJF seasons, respectively. (\textbf{E} to \textbf{F}) The scatter plots of the SST and sea surface salinity (SSS) climatology in the JJA season based on GODAS and \model's predictions at 12 months lead, respectively. The dashed contour lines denote potential density with a reference pressure of 0 dbar ($\sigma_0$ density, $\rm kg/m^3$). (\textbf{G} to \textbf{H}) As in (E) to (F), but for climatology in DJF season.} 
    \label{fig:tem_sal_vel_clim}
\end{figure}
\clearpage

\begin{figure}[p]
    \centering
    \includegraphics[width=0.99\textwidth]{pdf/temp_salt_acc_pattern.pdf}
    \caption{\textbf{Distributions of temporal Anomaly Correlation Coefficient (ACC) skills of \model~for temperature and salinity prediction.} (\textbf{A} to \textbf{C}) The ACC skills of \model~for sea surface temperature (SST) predictions at lead time of 6, 12, and 18 months, respectively. The black contour line is the dividing line where the prediction skill is equal to 0.5. (\textbf{D} to \textbf{F}), (\textbf{G} to \textbf{I}) and (\textbf{J} to \textbf{L}) As in (A) to (C) but for the prediction skill of temperature at a depth of 300m ($\rm T_{300}$), SSS, and salinity at a depth of 300m ($\rm S_{300}$), respectively.}
    \label{fig:temp_salt_acc_pattern}
\end{figure}
\clearpage

\begin{figure}[p]
    \centering
    \includegraphics[width=0.99\textwidth]{pdf/nino_pattern_index.pdf}
    \caption{\textbf{Skills of \model~in predicting ENSO.} (\textbf{A}) The DJF \nino~3.4 index based on GODAS and predictions of \model~at lead times of 3, 6, 12, and 24 months, respectively. The x-axis represents the winter season of each year, i.e., 2015 represents the period from December 2015 to February 2016. (\textbf{B}) Correlation skills of \nino~3.4 index prediction as a function of the lead month. The testing period is 1985-2018. (\textbf{C} to \textbf{D}) The 20\du C isotherm depth (d20) anomalies averaged over latitude (5\du S-5\du N) in the Pacific basin based on GODAS and \model~prediction at a lead time of 6 months respectively. The x-axis and y-axis denote the time and longitude, respectively. (\textbf{E} to \textbf{G}) The depth-longitude distribution of ACC skills of temperature prediction in the equatorial region at lead time of 6, 12, and 18 months, respectively.}
    \label{fig:nino_pattern_index}
\end{figure}
\clearpage

\begin{figure}[p]
    \centering
    \includegraphics[width=0.99\textwidth]{pdf/upper_ocean_mhw.pdf}
    \caption{\textbf{Forecast skills of upper ocean marine heatwaves.} (\textbf{A}) The distribution of the frequency of the occurrence of upper ocean marine heatwave events based on GODAS from 2010 to 2018. (\textbf{B} to \textbf{H}) The regionally average SEDI skill as a function of the lead time for seven selected regions with a relatively high frequency of heatwave occurrences. The skills are calculated based on MHW events for the period between 1985 and 2018. The higher SEDI denotes the better skill.}
    \label{fig:upper_ocean_mhw}
\end{figure}
\clearpage

\begin{figure}[p]
    \centering
    \includegraphics[width=0.999\textwidth]{pdf/decadal_prediction.pdf}
    \caption{\textbf{Decadal prediction of \model.} (\textbf{A}) The global mean SST based on GODAS and \model's prediction with initial field of December 1989, respectively. The solid line indicates the ensemble mean results while the shadow indicates the spread of \model's different members.
     (\textbf{B}) The global annual mean of SST anomalies based on GODAS and \model's predictions at lead times of 1, 3, and 5 years. The time range is selected as 1990-2019 to unify the forecasts for different lead times. (\textbf{C} to \textbf{E}) Correlation skills of the Pacific Decadal Oscillation (PDO), Interdecadal Pacific Oscillation (IPO), and Atlantic Multidecadal Oscillation (AMO) index forecast as a function of lead time, respectively. The testing period is also 1990-2019. The skills shown in the bar chart are calculated using annual mean (1y and 2y) and multi-year mean (2-5y etc.) data, with the x-axis representing lead years. In contrast, the skills depicted by the red line with square markers are based on monthly data, with the x-axis representing lead months (2m, 6m, etc.).}
    \label{fig:decadal_prediction}
\end{figure}



\clearpage 

%
\bibliography{ref} 
\bibliographystyle{scimag}


\section*{Acknowledgments}
We thank the climate modeling groups for producing and making available their model output, the Earth System Grid Federation (ESGF) for archiving the data and providing access, and the multiple funding agencies that support CMIP and ESGF. All CMIP data are available from the ESGF at https://esgf-node.llnl.gov/projects/esgf-llnl/. We would like to express our gratitude to the Research Support, IT, and Infrastructure team at the Shanghai Artificial Intelligence Laboratory for their valuable assistance in providing computation resources and network support throughout this research project.

\paragraph*{Funding:} J.-J L is supported by National Natural Science Foundation of China (Grant No. 42088101), National Key R\&D Program of China (Grant No. 2020YFA0608000). N.B. acknowledges funding by the Volkswagen Foundation and the European Union's Horizon Europe research and innovation programme under grant agreement No. 101137601.

\paragraph*{Author contributions:}
Conceptualization: Z.G., P.L., F.L., L.B., J.-J.L, and N.B. Methodology: Z.G., P.L., F.L., L.B., and J.-J.L. Investigation: Z.G., P.L., and F.L. Visualization: Z.G. Supervision: L.B., J.-J.L., and W.O. Writing-original draft: Z.G., P.L., F.L., L.B., J.-J.L, N.B., T.Y., T.I., S.C., and A.C.
\paragraph*{Competing interests:}
There are no competing interests to declare.
\paragraph*{Data and materials availability:}

CMIP6 database, https://esgf-node.llnl.gov/search/cmip6/; 
SODA, http://www.soda.umd.edu/; 
ORAS5, https://cds.climate.copernicus.eu/datasets/reanalysis-oras5;
GODAS, https://psl.noaa.gov/data/gridded/data.godas.html; 
SINTEX-F, http://www.jamstec. go.jp/aplinfo/sintexf/e/seasonal/outlook.html; 
NUIST-CFS1.2, https://icar.nuist.edu.cn/en/main.psp; 
NMME, https://iridl.ldeo.columbia.edu/SOURCES/.Models/.NMME/.


\subsection*{Supplementary materials}
Materials and Methods\\
Figs. S1 to S22\\
Tables S1 to S4\\
References \textit{(48-\arabic{enumiv})} 


\newpage


\renewcommand{\thefigure}{S\arabic{figure}}
\renewcommand{\thetable}{S\arabic{table}}
\renewcommand{\theequation}{S\arabic{equation}}
\renewcommand{\thepage}{S\arabic{page}}
\setcounter{figure}{0}
\setcounter{table}{0}
\setcounter{equation}{0}
\setcounter{page}{1} 


\begin{center}
\section*{Supplementary Materials for\\ \scititle}

Zijie~Guo,
Pumeng~Lyu,
Fenghua~Ling,\\
Lei Bai$^{\ast}$,
Jing-Jia Luo$^{\ast}$,
Niklas Boers,
Toshio~Yamagata,\\
Takeshi~Izumo,
Sophie~Cravatte,
Antonietta~Capotondi,
Wanli Ouyang\\ 
\small$^\ast$Corresponding authors. Email: bailei@pjlab.org.cn; luojj@nuist.edu.cn\\
\end{center}

\subsubsection*{This PDF file includes:}
Materials and Methods\\
Figures S1 to S22\\
Tables S1 to S4


\newpage



\subsection*{Materials and Methods}

\subsubsection*{Datasets}

Historical simulations from 20 models that participated in the CMIP6 are adopted for training (see Table~\ref{tab:cmip6}). And the data before 1980 from reanalysis datasets Simple Ocean Data Assimilation (SODA) \cite{carton2000simple} and Ocean Reanalysis System 5 (ORAS5) \cite{copernicus2021oras5} are used for model evaluation and parameters tuning. After training, we conduct independent testing of \model~with reanalysis data produced by NCEP GODAS from 1980 to 2018 (Table~\ref{tab:datasets_orca}).

For easier modeling, all data mentioned above are interpolated into the regular gird (63.5\textdegree S - 63.5\textdegree N, 0.5\textdegree - 359.5\textdegree E) with the resolution of 1\textdegree (128\texttimes 360 grid points). Also, the multi-level variables are interpolated into pre-defined depths, including 16 layers in depth (see Table~\ref{tab:vars}). Furthermore, the mean and standard deviation of each month in the SODA2 data (before 1980) are calculated for data normalization.

\subsubsection*{The hindcast of traditional dynamical models}
The North American Multi-Model Ensemble (NMME) is an experimental project, which was established in response to the U.S. National Academies’ recommendation to support regional climate forecasting and decision-making over intra-seasonal to interannual timescales. The project has been contributing model predictions from their hindcasts (dating back to the early 1980s) and real-time forecasts since August 2011. Each model consists of 6–28 ensemble members, and the forecasts are provided at lead times from 1 month to 11 months40 (https://iridl.ldeo.columbia.edu/SOURCES/.Models/.NMME/).

The SINTEX-F prediction system is built based on a fully coupled global ocean-atmosphere circulation model developed under the EU-Japan collaborative framework. This system has displayed high performance in predicting the tropical climate signals41. In particular, several ENSO events can be predicted at lead times of up to 2 years by this system14. The real-time predictions have been updated every month and made publicly available since 2006 (see http://www.jamstec.go.jp/aplinfo/sintexf/e/seasonal/outlook.html). 

The NUIST-CFS1.2 (Climate Forecast System of Nanjing University of Information Science and Technology) utilized in this paper is based on the fully coupled global ocean-atmosphere model SINTEX-F. This system employs an enhanced coupled SST-nudging scheme, utilizing the Ensemble Kalman Filter (EnKF) to assimilate multi-layer ocean data at the end of each month. The assimilated data encompasses full-field assimilation of observed sea surface temperature (SST), in-situ temperature and salinity profiles, as well as the assimilation of satellite-observed sea level anomalies (SLA). This approach enhances forecasting skills, particularly for subsurface condition, compared to the original SINTEX-F model (https://icar.nuist.edu.cn/en/main.psp).

\subsubsection*{Problem Formulation}

\model~is a skillful data-driven climate model that aims to predict the future ocean states based on the present ocean and atmosphere states. Formally, given the input ocean variables $O^{t}\in \mathbb{R}^{N_{\rm o} \times N_{\rm lat} \times N_{\rm lon}}$ and atmosphere variables (wind stresses) $A^{t}\in \mathbb{R}^{N_{\rm a} \times N_{\rm lat} \times N_{\rm lon}}$ at time $t$, lead time $\Delta t$, and ocean-land mask $\rm M \in \mathbb{R}^{N_{\rm lat} \times N_{\rm lon}}$, \model~is trained to produce the future ocean states $\hat{O}^{t+\Delta t}$, as shown in the Eq.~\ref{eq_problem_1},
\begin{equation}
    \hat{O}^{t+\Delta t}={\rm{\model}}(O^{t}, A^{t}, \Delta t, \rm M),
    \label{eq_problem_1}
\end{equation}
where $N_{\rm o}$ denotes the number of ocean predictands, $N_{\rm a}$ denotes the number of atmosphere variables, $N_{\rm lat}$ and $N_{\rm lon}$ denote the number of grids in latitude and longitude directions. Ocean areas in mask $\rm M$ have a value of 0, while land areas have a value of 1. In this study, we set $N_{\rm o}=66$, $N_{\rm a}=2$, $N_{\rm lat}=128$, $N_{\rm lon}=360$. In a single-step forecast, $\Delta t \in \{1, 2,.., K\}$, where $K$ is the maximum lead time that \model~can directly forecast. Since the model becomes more difficult to converge during training when $K$ gets larger, we set $K=6$ in this study. To generate forecasts with a lead time longer than $K$ months, we feed the model with the outputs $\hat{O}^{t+K}$ to get $\hat{O}^{t+K+\Delta t}$. By repeating this operation multiple times, we can generate forecasts for any lead time. The schematic diagram of the forecast process is shown in Fig.~\ref{fig:rollout_example}.

\subsubsection*{\model~Model}

Inspired by traditional ocean modeling frameworks, \model’s initial condition consists of multiple layers of ocean variables—temperature, salinity, and currents (see Table ~\ref{tab:vars} for complete modeling variables), while incorporating coastal lines as boundary conditions. Driving forces like wind stress on the ocean surface also influence ocean changes and are thus included. Since this work focuses solely on the upper 1000 meters of the ocean, bottom boundary conditions are not considered.
 In the modeling process, \model~independently encodes different variables to extract their own high-dimensional information. This information is then integrated in the fusion module to simulate the complex operations of the dynamic equations. 
 Finally, the decoder restores the calculated results to the spatial domain of each ocean variable, yielding next-step predictions. The overall architecture of \model~is illustrated in Fig.~\ref{fig:model}.

\noindent \textbf{Ocean Encoders and Decoders} consist of several ocean variable specific encoders and decoders. As \model~separately encodes and decodes each ocean variable, each variable corresponds to an encoder and decoder module. For each variable, the encoder module first patchifies it with a patch embedding module, which is a common operation in computer vision~\cite{dosovitskiy2020image}. Several consecutive local attention blocks then extract the high-level information. To capture features at different spatial scales while saving memory, down-sampling is conducted after every attention block except the last one following the implementation in Swin Transformer \cite{liu2021swin}. The decoder module has an inverse process like the encoder.

\noindent \textbf{Fusion Module} takes the merged ocean hidden states encoded by all encoder modules, atmosphere hidden states encoded by the atmosphere encoder, and the lead time $\Delta t$ as inputs. Inspired by the commonly used Rotary Position Embedding (RoPE) \cite{su2024roformer} in natural language processing, we transfer the position encoding to time encoding and propose the Time Rotary Fusion Module.

Let $\bm{x}_{\rm o}$, $\bm{x}_{\rm a}\in\mathbb{R}^{d}$ denotes the hidden states vector of a single grid in ocean and atmosphere hidden states respectively, where $d$ is hidden dims. We can get the rotated atmosphere vector $\bm{x}'_{\rm a} = \bm{R}^d_{\Theta,\Delta t}\bm{x}_{\rm a}$, where $\bm{R}^d_{\Theta,\Delta t}$ is the rotary matrix with pre-defined parameters $\Theta = \{\theta_i=10000^{-2(i-1)/d}, i \in [1, 2, ..., d/2]\}$ defined as follows:
\begin{equation}    
    \bm{R}^d_{\Theta,\Delta t} = 
    \begin{pmatrix}
    \cos{\theta_1\Delta t}& -\sin{\theta_1\Delta t}&0&0&\cdots&0&0\\
    \sin{\theta_1\Delta t}&\cos{\theta_1\Delta t}&0&0&\cdots&0&0 \\
    0&0&\cos{\theta_2\Delta t}& -\sin{\theta_2\Delta t}&\cdots&0&0\\
    0&0&\sin{\theta_2\Delta t}&\cos{\theta_2\Delta t}&\cdots&0&0 \\
    \vdots&\vdots&\vdots&\vdots&\ddots&\vdots&\vdots\\
    0&0&0&0&\cdots&\cos{\theta_{d/2}\Delta t}& -\sin{\theta_{d/2}\Delta t}\\
    0&0&0&0&\cdots&\sin{\theta_{d/2}\Delta t}&\cos{\theta_{d/2}\Delta t}
    \end{pmatrix}
    \label{eq_model_1}
\end{equation}
Then, the fused hidden vector $\bm{x}_{\rm f}$ can be calculated as follows:
\begin{equation}
    \bm{x}_{\rm f}=\bm{x}_{\rm o}+\langle \bm{x}_{\rm o},\bm{x}'_{\rm a}\rangle\cdot\bm{x}_{\rm a},
    \label{eq_model_2}
\end{equation}
where $\langle\cdot\rangle$ denotes the inner product. When $\Delta t$ gets larger, the relation between $\bm{x}_{\rm o}$ and $\bm{x}_{\rm a}$ gets weaker, quantified by $\langle \bm{x}_{\rm o},\bm{x}'_{\rm a}\rangle$ in Eq.~\ref{eq_model_2} (see also the original paper \cite{su2024roformer} for a detailed theoretical demonstration). With the Time Rotary Fusion Module, we can fuse the present input ocean hidden states and the initial atmosphere hidden states (present and initial conditions can be at a different time when doing autoregressive forecast), which avoids the challenging forecast of the wind stress \cite{zhou2023self} and improves \model's ability to manage the impact of atmospheric forcing on the ocean.

In order to further facilitate the model in extracting global information, position embedding \cite{dosovitskiy2020image} and several consecutive global attention blocks are added after fusion. 

\noindent \textbf{Attention Block} is mainly based on Swin Transformer \cite{liu2021swin}. It contains two successive sub-blocks. The first sub-block employs window-based multi-head self-attention (W-MSA) to capture the local information, while the subsequent sub-block uses shifted W-MSA (SW-MSA) for interactions across the windows. We introduce the ocean-land mask $\rm M$ to force the model to focus only on the ocean realm, similar to XiHe \cite{wang2024xihe}. To reduce the forecast steps and effectively simulate temporal differentiation or integration, just like the basic temporal ordinary differential equations (ODE), we replace the single MultiLayer Perceptron (MLP) module with a group of MLPs like the prevailing module Mixture-of-Experts (MoE) \cite{shazeer2017outrageously} used in natural language processing and use the lead time $\Delta t$ to index which MLP to use. This approach not only alleviates the accumulation of errors but allows the model to select different precursors at varying lead times, enhancing its capacity to capture seasonal cycles and effective simulation of temporal differentiation or integration, just like the basic temporal ordinary differential equations (ODE). The overall attention block is computed as follows:
\begin{align}
    &{{\hat{Z}}^{l}} = {\rm W\text{-}MSA}\Big( {{\rm LN}( {{{Z}^{l - 1}}} )}, \ {\rm M} \Big) + {Z}^{l - 1},\nonumber\\
    &{{Z}^l} = {\rm MLP}_{\Delta t} \Big( {{\rm LN}( {{{\hat{Z}}^{l}}} )} \Big) + {{\hat{Z}}^{l}},\nonumber\\
    &{{\hat{Z}}^{l+1}} = {\rm SW\text{-}MSA}\Big( {{\rm LN}( {{{Z}^{l}}} )}, \ {\rm M} \Big) + {Z}^{l}, \nonumber\\
    &{{Z}^{l+1}} = {\rm MLP}_{\Delta t}\Big( {{\rm LN}( {{{\hat{Z}}^{l+1}}} )}\Big) + {{\hat{Z}}^{l+1}},
    \label{eq_model_3}
\end{align}
where $\rm {LN}$ denotes the Layer Normalization \cite{ba2016layer}, $\rm M$ is the ocean-land mask, $\hat{Z}^{l}$ and $Z^l$ represent the output features of the (S)W-MSA and MLP for block $l$, respectively. The attention used inside the (S)W-MSA is defined as follows:
\begin{equation}
    {\rm Attention}(Q, K, V) = {\rm SoftMax}(QK^T/\sqrt{d}+{\rm M}*(-100))V,
    \label{eq_model_4}
\end{equation}
where $Q$, $K$, and $V$ represent the query, key, and values vectors, and d is the hidden dims. Instead of using the relative position bias \cite{liu2021swin}, we employ the RoPE to implicitly learn the relative position information. Moreover, the ocean-land mask $\rm M$ is added to the attention matrix to mask the land realm.

\noindent \textbf{Atmosphere Encoder} shares the same structure with the ocean encoder module but only uses one MLP in the attention sub-block as we do not forecast the atmosphere variables but employ the Time Rotary Fusion Module as mentioned above.

\subsubsection*{Training Details}
\model~is implemented with the Pytorch framework and trained with 4 Nvidia A100 GPUs within 12 hours using a total batch size of 32. The RMSE loss is used as the optimization target, which is defined as follows:
\begin{equation}
    {\mathcal{L}}_{\rm RMSE} = \sqrt{\frac{1}{N_{\rm o}\times N_{\rm lat}\times N_{\rm lon}}\sum_{v,i,j}{(\hat{O}^{t+\Delta t}_{v,i,j}-O^{t+\Delta t}_{v,i,j})^2}}.
    \label{eq_train_1}
\end{equation}

We adopt the Adam \cite{kingma2014adam} as the optimizer using the following parameters: $\beta_1=0.9$, $\beta_2=0.95$, $\epsilon = 1e^{-6}$ and ${\rm L}_2$ weight decay of $0.1$. The learning rate is warmed up with a ratio of $0.1$ to a maximum value of $2e^{-4}$, after which the cosine annealing is applied. 

\subsubsection*{Multi-member Prediction}

To reduce the uncertainty introduced by autoregression while further decreasing error accumulation, ten identical models are trained only with the difference of initial random seeds.

\subsubsection*{Evaluation Metrics}
\noindent \textbf{RMSE} (Root Mean Square Error) is a commonly used metric for evaluating how close a prediction is to the observation. Given the predicted values $\hat{O}^{t+\Delta t}$ and the observed values ${O}^{t+\Delta t}$, the RMSE of the prediction on a specific grid can be calculated as follows:
\begin{equation}
    {\rm RMSE}(v,i,j,\Delta t)=\sqrt{\frac{1}{T}\sum_{t}{(\hat{O}^{t+\Delta t}_{v,i,j}-O^{t+\Delta t}_{v,i,j})^2}},
    \label{eq_metrics_1}
\end{equation}
where $v$ denotes the specific variable or layer of multi-level variables, and $T$ is the number of test time points. To obtain the $\rm RMSE$ value for a region, we simply average the RMSE of each grid within the region.

\noindent \textbf{ACC} (Anomaly Correlation Coefficient) is a statistical measure used to evaluate how well a model is able to capture and reproduce the temporal phases of observed anomalies. The ACC of the prediction on a specific grid can be computed as follows:
\begin{equation}
    {\rm ACC}(v,i,j,\Delta t)=\frac{\sum_t(\hat{O}^{t+\Delta t}_{v,i,j}-\hat{C}^{{\rm m}_{t+\Delta t}}_{v,i,j})(O^{t+\Delta t}_{v,i,j}-C^{{\rm m}_{t+\Delta t}}_{v,i,j})}{\sqrt{\sum_t(\hat{O}^{t+\Delta t}_{v,i,j}-\hat{C}^{{\rm m}_{t+\Delta t}}_{v,i,j})^2\sum_t{(O^{t+\Delta t}_{v,i,j}-C^{{\rm m}_{t+\Delta t}}_{v,i,j})^2}}},
    \label{eq_metrics_2}
\end{equation}
where $\hat{C}$ and $C$ denote the forecast and observed climatology, ${\rm m}_{{t+\Delta t}}$ denotes the month corresponding to the time ${t+\Delta t}$. 
Note that, the $\rm ACC$ defined here is different from those in AI weather forecast models~\cite{bi2022pangu,lam2022graphcast,chen2023fengwu} which are the anomaly spatial correlation while we compute the temporal correlation. Additionally, The correlation skills refer to the Pearson correlation coefficient if not otherwise specified. 

\noindent \textbf{SEDI} (Symmetric Extremal Dependence Index) \cite{ferro2011extremal} is a measure for rare binary event forecast with several advantages, including non-degenerate, base-rate independent, asymptotically equitable, and so on. In this study, we use $\rm SEDI$ to assess the forecast performance of subsurface marine heatwaves. It is defined as follows:
\begin{equation}
    {\rm SEDI}=\frac{\log F-\log H-\log(1-F)+\log(1-H)}{\log F+\log H+\log(1-F)+\log(1-H)},
    \label{eq_metrics_3}
\end{equation}
where $H$ is the hit rate (true positive rate) and $F$ is the false alarm rate (false positive rate). SEDI has a range from $-1$ to $1$, and a higher value indicates better performance.

\subsubsection*{Definition of Upper Ocean MHWs}


In this study, upper ocean MHWs are defined with ocean heat content (OHC) following that of \cite{mcadam2023seasonal}. The OHC can be calculated as follows:
\begin{equation}
    {\rm OHC} = c_p \rho \int_{z_1=0{\rm m}}^{z_2=300{\rm m}}T(z)dz,
    \label{eq_definition_1}
\end{equation}
where $c_p$ is the specific heat capacity of seawater (3996 J/(kg.C)), $\rho$ is the density (1026 kg/$\rm m^3$), and $T(z)$ is the potential temperature at the depth $z$. The OHC anomalies are calculated first, then the MHW threshold corresponding to each month is calculated as the 90th percentile anomalies in a 3-month sliding window. For example, the threshold for January is the 90th percentile of all December to February OHC anomalies. Then, an MHW event is identified when the anomalies are greater than the 90th percentile of the corresponding month ~\cite{hobday2016hierarchical,jacox2022global,jacox2020thermal}. The climatology and 90th percentile are calculated based on the period of 1985-2018.

\subsubsection*{Calculation of Decadal Index}

\noindent \textbf{Pacific Decadal Oscillation} (PDO) is defined as the EOF's first mode of SST anomalies in the North Pacific Basin (north of 20°N). The SST anomaly is obtained by removing the corresponding climatology from the SST at each grid point and then removing the long-term trend of the climate. 
\noindent \textbf{Interdecadal Pacific Oscillation} (IPO) is represented by the TPI index \cite{henley2015tripole}. Three regions are defined first: (25\du N-45\du N, 140\du E-145\du W), (10\du S-10\du N, 170\du E-90\du W), and (50\du S-15\du S, 150\du E-160\du W). Then calculate the averaged SST anomalies in the three regions and get $\rm SSTA_1$, $\rm SSTA_2$, $\rm SSTA_3$. Finally, $\rm TPI=SSTA_2-(SSTA_1+SSTA_3)/2$. 

\noindent \textbf{Atlantic Multidecadal Oscillation} (AMO) is calculated as the averaged SST anomalies in the North Atlantic (north of 0\du) and the series is detrended. 







\newpage

\begin{figure}[p]
    \centering
    \includegraphics[width=0.99\textwidth]{pdf/model.pdf}
    \caption{\textbf{Architecture of the \model~model.} (\textbf{A}) \model~consists of several ocean encoders, an atmosphere encoder, a fusion module, and several ocean decoders (skipping connections between each encoder module and decoder module are not shown for brevity). The inputs include present ocean and atmosphere variables $O^{t}$ and $A^{t}$, ocean-land mask $\rm M$, and lead time $\Delta t$. The outputs are future ocean states $\hat{O}^{t+\Delta t}$. (\textbf{B}) Structure of the encoder module. (\textbf{C}) Structure of the fusion module. (\textbf{D}) Structure of the attention block. Local and global attention blocks share the same structure, but only with the difference in window size. (\textbf{E}) Structure of the decoder module.} 
    \label{fig:model}
\end{figure}
\clearpage

\begin{figure}[p]
    \centering
    \includegraphics[width=0.99\textwidth]{pdf/rollout_example.pdf}
    \caption{\textbf{Schematic diagram of forecast made by \model.} (\textbf{a}) \model~makes a single step forecast. $\Delta t$ indicates the lead time and $K$ is the maximum lead time that \model~can directly make a forecast without rolling out and is set to 6 in this study. The ocean-land mask is not shown in the figure for simplicity. Given the initial ocean variables $O^t$ and atmosphere condition $A^t$ at time $t$, and lead time $\Delta t$ ($\leq K$), \model~outputs the predicted ocean variables $\hat{O}^{t+\Delta t}$. (\textbf{b}) \model~makes a forecast autoregressively. Let $\Delta t=nK+r$ ($0\leq r<K$), \model~rolls out $n$ steps and outputs intermediate forecast $\hat{O}^{t+iK}$ at step $i$, where $i \in \{1,2,..,n\}$. At step $i$, lead time $\Delta \tau=iK$ is fed into the model to fuse the initial atmospheric conditions and the model outputs a forecast at a lead time of $K$ months (relative to intermediate input $\hat{O}^{t+(i-1)K}$). After $n$ rolling out steps, a final step is taken for the remaining $r$ months if $r>0$.}
    \label{fig:rollout_example}
\end{figure}
\clearpage

\begin{figure}[p]
    \centering
    \includegraphics[width=0.95\textwidth]{pdf/supp_temp_current_clim.pdf}
    \caption{\textbf{Climatological mean state of SST and SSC based on the period of 1985-2018.} (\textbf{a} to \textbf{j}) As in Fig.~\ref{fig:tem_sal_vel_clim} A-B but for GODAS in four seasons (MAM, JJA, SON, and DJF), \model~in MAM and SON seasons, and NUIST-CFS1.2 in four seasons, respectively.}
    \label{fig:supp_temp_current_clim}
\end{figure}
\clearpage

\begin{figure}[p]
    \centering
    \includegraphics[width=0.99\textwidth]{pdf/supp_temp_clim_bias.pdf}
    \caption{\textbf{The bias distribution of the mean states for SST during 1985-2018.} (\textbf{a} to \textbf{f}) As in Fig.~\ref{fig:tem_sal_vel_clim} C-D, but for \model~in MAM and SON seasons and NUIST-CFS1.2 in four seasons.}
    \label{fig:supp_temp_clim_bias}
\end{figure}
\clearpage

\begin{figure}[p]
    \centering
    \includegraphics[width=0.95\textwidth]{pdf/supp_temp_salt_clim.pdf}
    \caption{\textbf{Climatological mean state of SST and SSS during 1985-2018.} (\textbf{a} to \textbf{j}) As in Fig.~\ref{fig:tem_sal_vel_clim} E-H but for GODAS and \model~in MAM and SON seasons, and NUIST-CFS1.2 in four seasons, respectively.}
    \label{fig:supp_temp_salt_clim}
\end{figure}
\clearpage

\begin{figure}[p]
    \centering
    \includegraphics[width=0.99\textwidth]{pdf/supp_tem_sal_occurrence.pdf}
    \caption{\textbf{Density distribution of surface temperature-salinity points} (\textbf{a} to \textbf{d}) The distribution based on GODAS in different seasons during 1985-2018 respectively. The x and y axes are divided into 100 small intervals respectively, making a total of 100$\times$100 grids, and the number of points falling in each grid is counted. (\textbf{e} to \textbf{h}) and (\textbf{i} to \textbf{l}) As in (a) to (d), but for \model~and NUIST-CFS1.2 predictions, both at a lead time of 12 months.}
    \label{fig:supp_tem_sal_occurrence}
\end{figure}
\clearpage

\begin{figure}[p]
    \centering
    \includegraphics[width=0.99\textwidth]{pdf/supp_water_mass.pdf}
    \caption{\textbf{Temperature-Salinity plot in different water masses.} (\textbf{a} to \textbf{c}) The temperature-salinity plot based on the climatology of GODAS and predictions of \model and NUIST-CFS1.2 at a lead time of 12 months, respectively. The testing period is 1985-2018 and the climatology is averaged over all months. The points in different regions are displayed in different colors. The water masses include the Indian Ocean Equatorial Water (IEW), South Indian Ocean Central Water (SICW), Pacific Equatorial Water (PEW), Western North Pacific Central Water (WNPCW), Eastern North Pacific Central Water (ENPCW), Pacific Subarctic Upper Water (PSUW), Western South Pacific Central Water (WSPCW), Eastern South Pacific Central Water (ESPCW), Western North Atlantic Central Water (WNACW), Eastern North Atlantic Central Water (ENACW), and South Atlantic Central Water (SACW). The dashed contour lines denote potential density with a reference pressure of 0 dbar ($\sigma_0$ density, $\rm kg/m^3$). (\textbf{d} to \textbf{f}) As in (a) to (c), but colored by depth.}
    \label{fig:supp_water_mass}
\end{figure}
\clearpage

\begin{figure}[p]
    \centering
    \includegraphics[width=0.99\textwidth]{pdf/supp_clim_drift.pdf}
    \caption{\textbf{RMSE of climatological mean state of the period between 1985 and 2018.} (\textbf{a} to \textbf{f}) The RMSE of different seasons' climatological mean state for six variables, respectively. Different line colors denote the results of different models while different line styles denote different seasons. The RMSE of multi-layer variables like potential temp, current, and salinity is averaged over depth.}
    \label{fig:supp_clim_drift}
\end{figure}
\clearpage

\begin{figure}[p]
    \centering
    \includegraphics[width=0.99\textwidth]{pdf/supp_current_acc_pattern.pdf}
    \caption{\textbf{Distributions of ACC skills of \model~for the prediction of zonal and meridional current velocity.} (\textbf{a} to \textbf{f}) and (\textbf{g} to \textbf{l}) As in Fig.~\ref{fig:temp_salt_acc_pattern} (A) to (F), but for the zonal and meridional current velocity prediction, respectively.}
    \label{fig:supp_current_acc_pattern}
\end{figure}
\clearpage

\begin{figure}[p]
    \centering
    \includegraphics[width=0.99\textwidth]{pdf/supp_nuist_acc_pattern.pdf}
    \caption{\textbf{Distributions of ACC skills for NUIST-CFS1.2.} (\textbf{a} to \textbf{x}) As in Fig.~\ref{fig:temp_salt_acc_pattern} and Fig.~\ref{fig:supp_current_acc_pattern}, but for NUIST-CFS1.2.}
    \label{fig:supp_nuist_acc_pattern}
\end{figure}
\clearpage

\begin{figure}[p]
    \centering
    \includegraphics[width=0.99\textwidth]{pdf/supp_godas_std.pdf}
    \caption{\textbf{Distributions of the standard deviation (std) of anomalies based on GODAS during 1985-2018.} (\textbf{a} to \textbf{b}) The std of temperature anomalies at the surface and 300m, respectively. (\textbf{c} to \textbf{d}) As in (a) to (b), but for salinity anomalies.}
    \label{fig:supp_godas_std}
\end{figure}
\clearpage

\begin{figure}[p]
    \centering
    \includegraphics[width=0.99\textwidth]{pdf/supp_rmse_acc.pdf}
    \caption{\textbf{Forecast skills for the different variables.} (\textbf{a} to \textbf{f}) RMSE skills as a function of the lead time. The skills are calculated for anomalies and are averaged for all grids over the global ocean, with the test period spanning 1985–2018. (\textbf{g} to \textbf{l}) As in (a) to (f), but for ACC skills.}
    \label{fig:supp_rmse_acc}
\end{figure}
\clearpage

\begin{figure}[p]
    \centering
    \includegraphics[width=0.99\textwidth]{pdf/supp_vertical_rmse.pdf}
    \caption{\textbf{RMSE skills at each depth.} (\textbf{a} to \textbf{d}) The global averaged RMSE of potential temperature from \model, NUIST-CFS1.2, SINTEX-F, and persistence forecast, respectively. The x-axis and y-axis represent the forecast lead time and depth, respectively. (\textbf{e} to \textbf{p}) As in (a) to (d), but for the RMSE of zonal current, meridional current, and salinity prediction, respectively. All the skills are for anomalies.}
    \label{fig:supp_vertical_rmse}
\end{figure}
\clearpage

\begin{figure}[p]
    \centering
    \includegraphics[width=0.99\textwidth]{pdf/supp_vertical_acc.pdf}
    \caption{\textbf{ACC skills at each depth.} As in Fig.~\ref{fig:supp_vertical_rmse}, but for ACC skills}
    \label{fig:supp_vertical_acc}
\end{figure}
\clearpage

\begin{figure}[p]
    \centering
    \includegraphics[width=0.99\textwidth]{pdf/supp_ai_enso.pdf}
    \caption{\textbf{\nino~3.4 index forecast skills compared with other two specific deep-learning models.} (\textbf{a}) The correlation skill as a function of the lead time. \model~is compared with 3D-Geoformer~\cite{zhou2023self} and CNN~\cite{ham2019deep}. The testing period is unified to 1984-2017 for a fair comparison. (\textbf{b}) The performance of \model~for different target seasons. (\textbf{c}) The performance differences between \model~and 3D-Geoformer.}
    \label{fig:supp_ai_enso}
\end{figure}
\clearpage

\begin{figure}[p]
    \centering
    \includegraphics[width=0.99\textwidth]{pdf/supp_nino_ano_pattern.pdf}
    \caption{\textbf{The depth-time evolution of the
    mean temperature anomalies in \nino~3.4 region.} (\textbf{a} to \textbf{i}) The averaged anomalies based on GODAS and predictions of \model~and NUIST-CFS1.2 with lead times of 12, 18, and 24 months, respectively. The x-axis and y-axis denote the time and depth respectively.}
    \label{fig:supp_nino_ano_pattern}
\end{figure}
\clearpage

\begin{figure}[p]
    \centering
    \includegraphics[width=0.99\textwidth]{pdf/supp_d20.pdf}
    \caption{\textbf{Prediction of 20\du C isotherm depth.} (\textbf{a}) The 20\du C isotherm depth (d20) anomalies averaged over the equatorial Pacific basin (5\du S-5\du N, 120\du E-90\du W) based on GODAS and \model~prediction. (\textbf{b} to \textbf{f}) As in Fig.~\ref{fig:nino_pattern_index} (C) to (D) but for \model~prediction with lead times of 3, 12, and 24 months, respectively.}
    \label{fig:supp_d20}
\end{figure}
\clearpage

\begin{figure}[p]
    \centering
    \includegraphics[width=0.99\textwidth]{pdf/supp_indices.pdf}
    \caption{\textbf{Forecast skills of major climate indices.} (\textbf{a} to \textbf{f}) The forecast skills for six climate indices as a function of the lead time, including \nino~1+2 (10\du S-0\du, 90\du W-80\du W), \nino~3 (5\du S-5\du N, 150\du W-90\du W), \nino~4 (5\du S-5\du N, 160\du E-150\du W) SST anomaly, the Indian Ocean Dipole index (DMI)~\cite{saji1999dipole}, Atlantic \nino~index (ATL3)~\cite{zebiak1993air}, and Tropical Atlantic SST index (TASI)~\cite{chang1997decadal}.}
    \label{fig:supp_indices}
\end{figure}
\clearpage

\begin{figure}[p]
    \centering
    \includegraphics[width=0.99\textwidth]{pdf/supp_surface_mhw.pdf}
    \caption{\textbf{Forecast skills of surface MHWs.} (\textbf{a}) The distribution of the frequency of the occurrence of surface MHWs events based on GODAS. (\textbf{b}) Differences in the distribution of surface and upper ocean MHWs (Table~\ref{tab:mhw_region}). (\textbf{c} to \textbf{i}) The averaged SEDI as a function of the lead time for global and seven regions with a relatively high frequency of heatwave occurrences, the higher the better.}
    \label{fig:supp_surface_mhw}
\end{figure}
\clearpage

\begin{figure}[p]
    \centering
    \includegraphics[width=0.99\textwidth]{pdf/supp_decadal_diff_init.pdf}
    \caption{\textbf{The global mean SST based on GODAS and \model's prediction started from different initial fields.} (\textbf{a} to \textbf{e}) As in Fig.~\ref{fig:decadal_prediction} (A) but for the \model~forecasts initialized in different years.}
    \label{fig:supp_decadal_diff_init}
\end{figure}
\clearpage

\begin{figure}[p]
    \centering
    \includegraphics[width=0.99\textwidth]{pdf/supp_decadal_bias_depth.pdf}
    \caption{\textbf{The mean state bias of the decadal prediction of \model} (\textbf{a} to \textbf{c}) The bias of temperature, salinity, and current at the surface, respectively. The mean state is averaged over the 10-year rollout forecast initialized from December 1989 (1990-1999). (\textbf{d} to \textbf{f}), (\textbf{g} to \textbf{i}), (\textbf{j} to \textbf{l}), and (\textbf{m} to \textbf{o}) As in (a) to (c), but for the prediction bias at 100m, 300m, 500m, and 1000m depth, respectively.}
    \label{fig:supp_decadal_bias_depth}
\end{figure}
\clearpage

\begin{figure}[p]
    \centering
    \includegraphics[width=0.99\textwidth]{pdf/supp_godas_trend.pdf}
    \caption{\textbf{Trend distribution based on GODAS during 1990-2019.} (\textbf{a} to \textbf{c}) The multi-decadal trend of temperature, salinity, and current at surface, respectively. The trend is the slope coefficient of the least squares method and is calculated based on the annual mean data of GODAS (1990-2019). (\textbf{d} to \textbf{f}), (\textbf{g} to \textbf{i}), (\textbf{j} to \textbf{l}), and (\textbf{m} to \textbf{o}) As in (a) to (c), but for the trends at 100m, 300m, 500m, and 1000m depth, respectively.}
    \label{fig:supp_godas_trend}
\end{figure}
\clearpage



\begin{table}[p]
    \centering
    \caption{Variables and vertical levels modeled by \model. The input oceanic variables consist of two surface variables and four multi-layer variables with 16 layers. The input atmospheric variables only include two surface variables as initial condition and are not predicted by \model. }
    \begin{tabular}{ccc|c}
    \Xhline{1px}
        & Surface variables  & Multi-level variables & Vertical levels     \\ \hline
        \multirow{4}{*}{ocean}    & \multirow{2}{*}{surface temperature}   & potential temperature & \multirow{6}{*}{\makecell[c]{10, 15, 30, 50, \\ 75, 100, 125, 150, \\ 200, 250, 300, 400, \\ 500, 600, 800, 1000m}} \\
            &      & salinity     &           \\
            & \multirow{2}{*}{surface height above the geoid} & zonal velocity    &     \\
            &    & meridional velocity   &    \\ \cline{1-3}
    \multirow{2}{*}{atmosphere} & zonal wind stress                               & \multirow{2}{*}{/}    &                     \\
            & meridional wind stress   &   &  \\
    \Xhline{0.8px}
    \label{tab:vars}
\end{tabular}
\end{table}
\clearpage

\begin{table}[p]
  \centering
    \caption{CMIP6 models used for training.}
    \renewcommand\arraystretch{0.85}
  \begin{tabularx}{\textwidth}{>{\hsize=0.3\hsize\centering\arraybackslash}X>{\hsize=0.6\hsize\centering\arraybackslash}X}
    \toprule
    \textbf{Source ID} & \textbf{Institution} \\
    \midrule
    BCC-CSM2-MR & Beijing Climate Center  \\
    CAS-ESM2-0 & Chinese Academy of Sciences \\
    CIESM & Department of Earth System Science, Tsinghua University \\
    CMCC-CM2-HR4 & Fondazione Centro Euro-Mediterraneo sui Cambiamenti Climatici \\
    CMCC-CM2-SR5 & Fondazione Centro Euro-Mediterraneo sui Cambiamenti Climatici \\
    CMCC-ESM2 & Fondazione Centro Euro-Mediterraneo sui Cambiamenti Climatici \\
    E3SM-1-0 & E3SM-Project \\
    E3SM-1-1 & E3SM-Project \\
    E3SM-2-0 & E3SM-Project \\
    EC-Earth3 & EC-Earth Consortium \\
    EC-Earth3-Veg & EC-Earth Consortium \\
    FGOALS-f3-L & Chinese Academy of Sciences \\
    FIO-ESM-2-0 & First Institute of Oceanography, Ministry of Natural Resources / Qingdao National Laboratory for Marine Science and Technology \\
    HadGEM3-GC31-MM & Met Office Hadley Centre \\
    INM-CM4-8 & Institute for Numerical Mathematics, Russian Academy of Science \\
    INM-CM5-0 & Institute for Numerical Mathematics, Russian Academy of Science \\
    MPI-ESM1-2-HR & Max Planck Institute for Meteorology / Deutscher Wetterdienst / Deutsches Klimarechenzentrum \\
    MRI-ESM2-0 & Meteorological Research Institute \\
    SAM0-UNICON & Seoul National University \\
    TaiESM1 & Research Center for Environmental Changes, Academia Sinica \\
    \bottomrule
  \end{tabularx}
  \label{tab:cmip6}
\end{table}
\clearpage

\begin{table}[p]
    \centering
    \caption{The range of each region considered for MHWs forecast. The North Central Pacific and Central Atlantic regions are only used for surface MHWs.}
    \begin{tabular}{ccc}
    \toprule
    \textbf{Region} & \textbf{Range}\\         
    \midrule
     Northeast Pacific & (35\du N-60\du N, 160\du W-125\du W) \\
     North Central Pacific & (5\du N-30\du N, 140\du E-100\du W)  \\
     South Pacific & (60\du S-20\du S, 165\du E-150\du W)  \\
    Northwest Atlantic & (20\du N-40\du N, 80\du W-30\du W) \\
    Central Atlantic & (10\du S-15\du N, 60\du W-10\du W)  \\
    Southwest Atlantic & (50\du S-30\du S, 60\du W-10\du W)  \\
     South Indian Ocean & (40\du S-15\du S, 50\du E-100\du E) \\
    Mediterranean Sea & (25\du N-50\du N, 0\du E-40\du E) \\
    Bering Sea & (45\du N-60\du N, 170\du E-170\du W) \\
    \bottomrule
    \end{tabular}
    \label{tab:mhw_region}
\end{table}
\clearpage

\begin{table}[p]
    \centering
    \caption{Datasets used for training, validation, and testing.}
    \begin{tabular}{ccc}
    \toprule
     & \textbf{Data} & \textbf{Period} \\
    \midrule
    Training  & CMIP6 & 1850-2014 \\ \cmidrule{2-3}
    \multirow{2}{*}{Validation} & SODA2 & 1871-1979 \\ 
                              & ORAS5 & 1958-1979 \\ \cmidrule{2-3}
    Testing  &  GODAS &  1980-2019 \\ 
    \bottomrule
    \end{tabular}
    \label{tab:datasets_orca}
\end{table}
\clearpage


\clearpage 





\end{document}